# GPM Ground Validation
# Basic Radar Products and
# Implications for Observation Strategies

## Report from Working Group 4b
## 1st International GPM GV Requirements Workshop
## 4 – 7 November 2003; Abingdon, U.K.

Sandra Yuter, Jarmo Koistinen, Sabatino Di Michelle,
Martin Hagen, Anthony Illingworth, Shuji Shimizu, and Dave Wolff

## Summary

Recommendations are made for NASA/JAXA Global Precipitation Measurement (GPM) satellite Ground Validation (GV) program. This report details recommended GV site local radar products based on data from surface-based scanning radars including S-band, C-band, and X-band polarimetric and non-polarimetric radars. Three general categories of products are described: text products summarizing information on the statistical characteristics of the radar data and derived parameters, 2D products providing maps of the horizontal variability of near surface radar observed and derived parameters, and 3D products describing volumetric echo structure. Regional composites could include products based on several of the 2D and 3D single radar products. Several types of time-integrated 2D and 3D products are also recommended. A brief discussion of useful ancillary data from other sources and remaining challenges concludes the report.

## 1. Background

The recommended GV radar products address the scientific objectives defined at the GPM GV Working Group meeting in Seattle, Washington in February 2002 (Yuter et al. 2002). These objectives are:

a) Determination of minimum detectable surface precipitation rate.
b) Classification of precipitation in (x,z) and (x,y) dimensions into hydrometeor categories such as rain, snow, mixed, and graupel/hail.
c) Classification of the three-dimensional precipitation structure.
d) Determination of spatial pattern of surface precipitation intensity
e) Quantitative estimation of surface precipitation rate.
f) Description of errors associated with each of the above items (a-e).

During the October 2003 Precipitation Measurement Missions meeting in Greenbelt, Maryland, tasks related to the evaluation of the satellite estimated vertical profile of latent heating were added to the GPM GV responsibilities. The specific objectives for latent heating validation are still being defined but include collection of volumetric horizontal divergence data and classification of radar echo into convective and stratiform precipitation components.

The working group focused on potential GV products associated with surface-based scanning radars including S-band, C-band or X-band Doppler polarimetric and non-polarimetric radars. Based on presentations at the meeting, potential GV partner radars include a mixture of operational radars with fixed scan strategies and research radars with flexible scan strategies. It is anticipated that logistical, geographical, and operational constraints will limit partner radars to data collection for a subset of the recommended products. The full list of recommended products in this document represents a superset of products from a variety of radar types and locations.

## 2. General Recommendations

These recommendations represent the consensus of the working group.

1) It is vital that information on uncertainties be routinely included with every GV observed and derived product.

2) Polar coordinate data from participating radars will be archived and available for use by all GV partners (see Section 3).

3) Cartesian products are easier to compare to satellite data than polar coordinate data since range dependencies can be minimized when a Cartesian grid is appropriately scaled for the radar characteristics and a maximum product range. Since some interpolation schemes can introduce artifacts and degrade information, a common high quality objective interpolation methodology needs to be agreed upon and adapted to insure quality and consistency among Cartesian products from different sites.

4) For all GPM applications, snow accumulation should be expressed in units of equivalent liquid per unit time. Snow accumulation depths vary with crystal shape and temperature and hence are difficult to compare among sites and storms.

5) Observed products such as radar reflectivity should have non-meteorological echoes identified with a bad quality flag and confidence level. Non-meteorological echoes are to be removed before derived products are calculated.

## 3. Archival of Original Radar Measurements

Polar coordinate radar data represent a key data set for GV. All other products are derived (and reprocessed) from the originally recorded polar data. It is vital to the success of GV that the originally recorded radar data be archived and accessible. The archived format should either be a public domain format or one that is readily transferable to a public domain format. The archived polar coordinate data must preserve information on the azimuth angle relative to true north, the elevation angle, and the range gate spacing.

To be of value for error characterization products utilized by operational users, GV radar data needs to be available soon after data collection. Ideally, partner radar data should be deposited at the designated GV archive within 48 hours of data collection. It is recognized that some radar data sets may not be available within 48 hours. These data sets will be valuable for detailed analysis and climatological applications which have less stringent latency requirements.

Quality control (QC) should be applied to the polar coordinate data prior to processing of derived products. Local knowledge should be applied as much as possible in the removal of non-meteorological echoes especially for QC parameters which vary with space and time. GPM GV may find it useful to apply a second round of QC if it is capable of detecting additional non-meteorological echoes or if the automated national system does not include advanced QC techniques.

Distinguishing pixels with no precipitation from those with lost precipitation signal will likely be a demanding task. The latter category includes beam blocking due to mountains which can vary with the thermodynamic profile and radar beam overshooting of shallow precipitation at long ranges which can vary storm to storm. Individual quality factors at each pixel could be summed into a general quality class flag.

## 4. Text Products

The purpose of text products is to provide an easy to use summary of radar observed and derived products. Likely applications of text products are to identify times of interest for detailed study within the GV data sets, and for comparison of basic statistics among the GV sites, GPM satellites, and regional model output.

Table 1. Recommended Text Products. By definition all products contain information on the uncertainty of their constituent variables.

| Name | Description |
| --- | --- |
| Precip Area above Z thresholds | X km$^2$ of precip within radar domain of Y km$^2$ for Z thresholds of 0, 10, 20 dBZ |
| Surface precipitation types present | Rain, snow, mixed, graupel/hail |
| Presence of distinct melting layer | Yes or No in volume |
| Rain layer depth | km – average, standard deviation, min, max within volume |
| Ice layer depths for Z thresholds | km - average, standard deviation, min, max within volume for Z thresholds of 0, 10, 20 dBZ |
| Echo top height for Z thresholds | km – average, standard deviation for Z thresholds of 0, 10, 20 dBZ |
| Attenuation correction (X- and C-band radars) | Whether attenuation correction is applied and some information on its application. |
| Coincident GPM core and constellation satellite overpasses | Time, distance to nadir, ascending or descending and name of the satellite |
| Z, (ZDR) calibration | Offset and its uncertainty versus recorded data |

*Notes on text products*

Suggested Z thresholds include values below and above expected GPM core satellite DPR sensitivity.

Echo top height is dependent on radar sensitivity and scan strategy which varies from radar to radar. Hence echo top height statistics will be difficult to compare among sites.

## 5. 2D Near-surface Scan Products

The purpose of 2D near-surface scan products is to document the horizontal variability of observed and derived parameters. 2D products are based on a single low-level elevation angle PPI scan ($\leq 1°$) These products are distinguished from 3D products which require volume scans, i.e. a set of PPIs at different elevation angles. Applications of 2D products include comparison with a wide range of satellite intermediate and final map products.

All the products on the table below are recommended to be on an objectively interpolated Cartesian grid (Trapp and Doswell 2000). Since radar characteristics such as beam width and the maximum usable product range vary among radars and precipitation vertical structures, the optimal Cartesian grid resolution and size may be radar and seasonally specific.

When multiple coordinated radars are available within a region, some of the 2D map products below may be able to be produced as regional composites (Section 8).

As part of GV site documentation, a detailed map is needed of the surface background types in terms of water, land, and potentially land-surface type (urban, forest, grassland etc) for each pixel in the 2D Cartesian grid.

Table 2. Recommended 2D Products. QC- stands for quality controlled variables. By definition all products contain information on the uncertainty of their constituent variables.

| Name | Description |
|---|---|
| QC-Z long range surveillance scan | Cartesian map of observed reflectivity based on low PRF surveillance scan at low elevation angle |
| QC-observed radar variables within lowest tilt of volume scan | Cartesian map of each of the observed parameters which serve as input for derived 2D products. |
| Location of non-meteorological echoes | Diagnostic map. Non-meteorological echo tagged with bad quality flag and confidence. May be incorporated into QC-Z and QC-observed products. |
| Attenuation correction (X- and C-band radars) | Diagnostic PPI scan showing attenuation correction values in dB at each polar coordinate pixel. |
| Range where radar beam overshoots precipitation. | Diagnostic map of usable radar range for each volume. Of particular interest for shallow precipitation. |
| Precipitation echo locations | Derived Cartesian map. Existence of precipitating echo for various Z thresholds (0, 10, 20 dBZ). For some applications mm/h thresholds may be useful. |
| Surface precipitation hydrometeor type | Derived Cartesian map of estimated surface values. Including rain, snow, mixed rain and snow, graupel/hail. Of particular interest in midlatitudes. |
| Convective and stratiform precipitation | Derived Cartesian map. Objective classification of radar echo regions. Of particular interest in the tropics for use in partitioning latent heating. |
| R derived from Z | Derived Cartesian map of estimated surface precipitation intensity (rainrate or snowfall) based on reflectivity, scan geometry, and storm structure without use of polarimetric |

|  | parameters. The vertical profile correction applied to each pixel should be included as a field in this product. |
|---|---|
| Hybrid R | As above except including available polarimetric variables as appropriate. |

*Notes on 2D products:*

2D maps include products representing both data at the height of the radar beam and estimates of the real ground/sea surface value derived from measurements at the height of the radar beam which is always above the surface. The precipitation type and rain rate will be compared to surface-based in situ instruments. These categories of map products are by definition estimates of the surface values which take into account the vertical profile of the precipitation.

Overlaying the surface precipitation intensity and surface precipitation hydrometeor type maps will distinguish snow rates from rain rates.

A -10 dBZ threshold within 50 km range of the radar would be useful for precipitation echo locations in snow.

Depending on how the GV convective/stratiform classification algorithm is modified for GPM it may become a 3D product. The current TRMM GV convective stratiform product is a 2 km x 2 km horizontal resolution Cartesian grid to 150 km range based upon information in the lowest tilt of the radar volume scan (Steiner et al. 1995). The algorithm parameters are tuned for a particular precipitation climatology and radar with volumetric data (Yuter and Houze 1997). The current TRMM PR convective stratiform algorithm utilizes a combination of horizontal (5 km x 5 km) and vertical (250 m at nadir) precipitation echo information (Awaka et al. 1998).

The maximum range that hybrid techniques for estimating R can be successfully applied is a subject of active research within the radar community. Within the literature, these techniques are often restricted in range compared to non-polarimetric methods.

## 6. 3D Volume Scan Products

3D products require volume scans containing several elevation angles. These products document the 3D variability of observed and derived parameters. Some of these products are 2D maps derived from volumetric data. The working group decided it made more sense to group products by type of radar scan needed to produce them as compared to their output format. For example, echo top height is a 2D map but 3D data are required to derive it so it is classified as a 3D product.

Primary applications are comparison to 3D observed and derived products from the DPR and for assessment of physical assumptions regarding the 3D storm structure used in estimating surface precipitation from satellite passive microwave observations. Divergence products are for evaluation of satellite-derived latent heating estimates.

With the exception of the RHI product (see notes below) and the VVP and VAD products, the balance of products in the Table 3 below are recommended to based on data objectively interpolated to a 3D Cartesian grid. See comments on 2D products (Section 5) regarding grid resolutions and sizes.

Table 3. Recommended 3D Products. QC- stands for quality controlled variables. By definition all products contain information on the uncertainty of their constituent variables.

| Name | Description |
| --- | --- |
| RHI along satellite track during GPM core satellite overpasses with precipitation. | For comparison to DPR observations, and satellite-derived attenuation correction, and $D_o$. |
| QC-observed radar variables within volume | 3D Cartesian volume of observed parameters which serve as input for derived Cartesian 3D products. |
| Location of non-meteorological echoes | Diagnostic volume. Non-meteorological echo tagged with bad quality flag and confidence. May be incorporated into QC-observed products. |
| Attenuation correction (X- and C-band radars) | Diagnostic PPI volume showing attenuation correction values in dB at each polar coordinate pixel. |
| Echo top | Derived Cartesian map of maximum height of precipitating echo for various Z thresholds (0, 10, 20 dBZ). |
| Rain layer height | Derived Cartesian map. Expected to vary across baroclinic fronts. |
| Hydrometeor classification | Derived Cartesian 3D volume. Echo volume pixels classified by hydrometeor type, requires polarimetric parameters |
| Vertically integrated LWC and IWC | Derived Cartesian map of vertically integrated values based on Z or polarimetric parameters. |
| Velocity Volume Processing (VVP)-yields profiles of horizontal winds, divergence and Z | Derived profiles based on cylindrical volume of radial velocity and Z data within 30-40 range of radar. |
| Velocity Azimuth Display (VAD)- horizontal wind and divergence profiles | Derived profile based on conical volume of radial velocity data within precipitation echo. |
| CFADs of Z by categories | Derived array of joint frequency distribution with height of Z for categories of precipitation echo within 3D Cartesian volume. |

| Vertical profiles of Z by categories | Derived from precipitation echo within 3D Cartesian volume. |
|---|---|
| CFADs of horizontal divergence by categories for multiple Doppler or bistatic sites. | Derived array of joint frequency distribution with height of divergence for categories of precipitation echo within 3D Cartesian volume. |
| Vertical profiles of horizontal divergence by categories for multiple Doppler or bistatic sites. | Derived from precipitation echo regions within within 3D Cartesian volume. |

*Notes on 3D products*

The working group makes a strong recommendation to obtain RHIs parallel to the GPM core satellite track within the DPR $K_u$-band swath. To be of most value, the largest dimension of the GV radar's effective beamwidth should not exceed the ground resolution of the DPR instrument. Only a subset of GV partner radars will have the operational flexibility to obtain these RHIs. Applications of this RHI product may require both polar coordinate and objectively interpolated Cartesian versions.

Echo top height, see notes in Section 4. Also a -10 dBZ threshold would be useful within 50 km range of radars with sufficient sensitivity.

Rain layer height is similar but not identical to bright band height. Rain layer height is a parameter used in radiative transfer calculations.

Contoured frequency by altitude diagrams (CFADs, Yuter and Houze 1995) of Z by categories are a current TRMM GV product. Their purpose is to provide concise information on the frequency distribution with altitude of Z such as skewedness, modes, min, and max etc.

Categories of precipitation for the profiles and CFADs include the combinations of total, convective, and stratiform precipitation components with the surface types of land, ocean, coastal, and all used for TRMM. For GPM, additional categories are added for midlatitude sites related to surface precipitation types such as rain, snow, mixed, and graupel/hail.

The quality of the VVP (Waldteufel and Corbin, 1979; modified by Koscielny et al., 1982) and VAD (Lhermitte and Atlas, 1961; Browning and Wexler, 1968) products is a function of the volume scan strategy, i.e. the number and spacing of elevation angles and the maximum elevation angle, and how well the assumption of horizontal homogeneity holds within the analyzed volume. Stratiform precipitation, which is more uniform in the horizontal, usually yields higher quality VVP and VAD output than convective precipitation. In practice, these products are usually derived for a subset of the scanned volume less than 50 km range from the radar.

## 7. Time-integrated Products

The working group recommends the use of only two accumulation time periods which can be combined by users into a variety of customizable time scales such as 5-day, calendar months, 30-day periods, seasonal, etc. The two recommended time scales are:

- 24 hours (0000 UTC-2359 UTC)
- Storm duration

Storm duration may be difficult to define precisely. We suggest guidelines be developed for GPM GV on this topic. Based on working group discussions and the current TRMM products the time-integrated products in Table 4 are suggested.

Table 4. Recommended time-integrated products for both 24 hour and storm duration time periods. See discussion in Section 6 on precipitation categories. By definition all products contain information on the uncertainty of their constituent variables.

| Name |
| --- |
| Rainfall accumulation maps in units of mm height per $m^2$ area |
| Snowfall accumulation maps in units of equivalent liquid mm height per $m^2$ area |
| Accumulated vertical profiles of Z by precipitation categories |
| Accumulated CFADs of Z by precipitation categories |

Included with all time integrated products should be information on the time interval between the radar scans used to compute the product. The time interval between low level scans should ideally be $\leq 5$ min. For each time-integrated product, information on data gaps and the associated confidence level are also needed. For example, for days with no data gaps or if the radar was taken down during a non-precipitating period for scheduled maintenance, the confidence that the 24 hour total represents the actual accumulation would be high. In contrast, if the radar broke down in the middle of a storm, the confidence would be low. Note that confidence level in this context relates to temporal sampling gaps. Information on the total uncertainty of the time-integrated products based on their sampling interval and on the uncertainties in the instantaneous precipitation rate and Z products is also needed.

## 8. Regional Composite Products

Regional composite products based on data from multiple coordinated radars (Michelson et al. 1999, Raschke et al. 2001, Koistinen and Michelson 2002) have the benefits large area, overlapping coverage, and often enhanced quality compared to single radar products. Since these data are likely to be processed and reprocessed for multiple applications, data arrays are needed for GPM GV applications. Potential regional composite products may be able to be adapted from a subset of the 2D and 3D products discussed in Sections 5 and 6.

## 9. Ancillary Data

The working group briefly discussed several other types of surface-based measurements that would add significant value to the weather radar observations. These are:

- Ceilometer cloud base heights
- Vertical profiles of water and ice content from cloud radars.

Additionally, NWP outputs for the GV radar domain including the hourly thermodynamic profile above each site, 3D winds, and cloud and rain parameters would be very useful to have in a form that facilitates comparison with the radar products. It is recommended that the NWP outputs over the GV sites data be developed as a GV product and archived with the local radar site products.

## 10. Some Remaining Challenges

The successful validation of satellite-derived snow estimates requires progress on several challenges in snow measurement from the ground. Current methods for estimation of snow rate from observed radar reflectivity are widely considered to be unreliable. A research focus on this problem is needed in order to characterize and potentially reduce uncertainties to acceptable levels for GPM applications. Another important challenge is the validation of light snow rates which is difficult with current instruments. Some new instruments including the DRI/NCAR hotplate and weighing gauges by Geonor and OTT need to be tested for GPM applications and their uncertainties characterized under a range of conditions.

In the course of writing up this report, several issues arose which will need to be addressed at future meetings:

Given differences in radar beam widths and maximum usable range, to what degree should Cartesian product grid resolutions and sizes be standardized among partner radars.

Where is the responsibility for the removal of non-meteorological echo in the observed radar parameters? Is it to be flagged at the local site before the data is sent on to the GV archive or should QC be performed on the archived data prior to product processing? It may be difficult to require a standardized QC methodology to be performed by operational radars.

**References**

Awaka, J., T. Iguchi, and K. Okamoto, 1998: Early results on rain type classification by the Tropical Rainfall Measuring Mission (TRMM) precipitation radar. Proc.

URSI-F Symp. on Wave Propagation and Remote Sensing. Aveiro, Portugal, 143-146.

Browning, K.A. and R. Wexler (1968): The determination of kinematic properties of a wind field using Doppler radar. *J. Appl. Meteor.*, **7**, 105 - 113.

Koistinen, J., and D. B. Michelson, 2002: BALTEX weather radar-based precipitation products and their accuracies. *Boreal Env. Res.*, 7, 253-263.

Koscielny, A.J., R.J. Doviak and R.Rabin (1982): Statistical considerations in the estimation of divergence from single Doppler radar and application to prestorm boundary layer observations. *J. Appl. Meteorol.*, **21**, 197 - 210.

Lhermitte, R.M. and D.A. Atlas (1961): Precipitation motion by pulse Doppler radar. Proc. 9th Weather Radar Conf. Boston, Amer. Meteor. Soc., Boston, 218 - 223.

Michelson, D. B., T. Anderson, C. G. Collier, Z. Dziewit, J. Koistinen, S. Overgaard, J. Riedl, and V. Zhukov, 1999: The international radar network for the Baltic Sea Experiment. *Preprints, 29$^{th}$ Radar Meteorology Conference*, 12-16 July 1999, Montreal, Canada, AMS, Boston, 317-320.

Raschke E., J. Meywerk, K. Warrach, U. Andrae, S. Bergström, F. Beyrich, F. Bosveld, K. Bumke, C. Fortelius, L. P. Graham, S. –E. Gryning, S. Halldin, L . Hasse, M. Heikinheimo, H. –J., Isemer, D. Jacob, I. Jauja, K. –G. Karlsson, S. Keevallik, J. Koistinen, A. van Lammeren, U. Laas, J. Launiainen, A. Lehmann, B. Liljebladh, M. Lobmeyer, W. Matthäus, T. Mengelkamp, D. B. Michelson, J. Napiorkowski, A. Omstedt, J. Piechura, B. Rockel, F. Rubel, E. Ruprecht, A. –S. Smedman, and A. Stigebrandt, 2001: The Baltic Sea Experiment (BALTEX): A European contribution to the investigation of the energy and water cycle over a large drainage basin. *Bull. Amer. Meteor. Soc.*, **82**, 2389-2413.

Steiner, M., R. A. Houze, Jr., and S. E. Yuter, 1995: Climatological characterization of three-dimensional storm structure from operational radar and rain gauge data. *J. Appl. Meteor.*, **34**, 1978-2007.

Trapp, R. J., and C. A. Doswell III, 2000: Radar data objective analysis. *J. Atmos. Ocean. Tech.*, **17**, 105-120.

Waldteufel, P. and H. Corbin (1979): On the analysis of single Doppler radar data. *J. Appl. Meteor.*, **18**, 532 - 542.

Yuter S. E., and R. A. Houze, Jr., 1995: Three-dimensional kinematic and microphysical evolution of Florida cumulonimbus, Part II: Frequency distributions of vertical velocity, reflectivity, and differential reflectivity. *Mon. Wea. Rev.*, **123**, 1941-1963.

Yuter, S. E., and R. A. Houze, Jr., 1997: Measurements of raindrop size distributions over the Pacific warm pool and implications for Z-R relations. J. Appl. Meteor., **36**, 847-867.

Yuter, S., R. Houze, V. Chandrasekar, E. Foufoula-Georgiou, M. Hagen, R. Johnson, D. Kingsmill, R. Lawrence, F. Marks, S. Rutledge, and J. Weinman, 2002: GPM Draft Science Implementation Plan Ground Validation Chapter. accessible at: www.arxiv.org, arXiv:physics/0211095, 22 pp.